\documentclass[11pt,epsf,amstex]{article}
\usepackage [dvips]{graphicx}
\usepackage{amsmath}
\usepackage{times}
\usepackage{psfig}
\usepackage{epsfig}
\usepackage{amssymb}

\newcommand{\dd}{\text{d}}

\newcommand{\ee}{\text{e}}

\newcommand{\p}{\partial}
\newcommand{\bx}{\text{\bf x}}
\newcommand{\br}{\text{\bf r}}
\newcommand{\by}{\text{\bf y}}

\newcommand{\eps}{\varepsilon}
\newcommand{\bk}{\text{\bf k}}

\newcommand{\bq}{\text{\bf q}}

\begin{document}
\newenvironment{dessins}{{\bf Figures}}{}
\begin{center}
{\huge{\bf Trapping of a random walk\\
by diffusing traps}}
\end{center}
\vspace{2cm}
\begin{center}

Fr\'ed\'eric van Wijland
\end{center}

\noindent {\small P\^ole Mati\`ere et Syst\`emes Complexes (CNRS FR2438, Universit\'e de Paris VII) and Laboratoire de Physique Th\'eorique, CNRS
UMR8627, Universit\'e de Paris-Sud,
91405 Orsay cedex, France.}\\\\

\begin{center}{\bf Abstract}\\
\end{center}
{\small We present a systematic analytical approach to  the trapping of a random walk by a finite density
$\rho$ of diffusing traps in arbitrary dimension $d$. We
confirm the phenomenologically predicted $\ee^{-c_d \rho t^{d/2}}$ time decay of the survival
probability, and compute the dimension dependent constant $c_d$ to leading order within an $\eps=2-d$
expansion.}

\vskip 2cm


\newpage

\section{Introduction}
\subsection{Motivations}
It has been over thirty years that  the
trapping of a random walk in a medium with absorbing traps is the focus of
physicists'attention. The reason is
twofold. First and more importantly the experimental relevance of modeling
the diffusion of exciton has led to a compact formulation of the problem in
terms of trapping of random walks. The second motivation is of theoretical
nature. 
Much is known on the static traps case. Connections between trapping and Lifschitz
tails in disordered electronic systems~\cite{tao}, statistics of rare
events~\cite{varadhan} and Brownian motion
theory have been established. That the problem bridges to other areas of physics
stems from the ubiquity of random walks and their applications.

Here we will concentrate our study on the diffusion of a random walker in the presence of
{\it diffusing traps}. The configuration of the traps evolves in time, instead of
remaining frozen in its initial distribution. Despite the simplicity of its
formulation, this consitutes a genuine many-body problem in which an infinite
number of degrees of freedom are coupled: the positions of the traps relative to
that of the random walker are correlated.


Let us provide a more precise formulation of the problem we propose to
investigate. We are interested in the problem of a tagged walker (hereafter christened as 
``the walker'') evolving in a medium in
which a finite density $\rho$ of diffusing traps is present. Both the traps and
the tagged walker have the same diffusion constant. Here we have set the
diffusion constant to $\frac{1}{2}$ for
aesthetic reasons. However it would be of interest to investigate the diffusion
constants ratio dependence, since, as shown in \cite{russes}, at sufficiently
low trap mobility, there exists a rich crossover regime between static and mobile
traps. When the walker and a trap meet on the same site then, between $t$ and
$t+\dd t$, the walker has a probability $\beta\dd t$ to
die.\\

The static traps case
was ``solved'' some twenty years ago using field-theoretic methods based on instanton
calculus~\cite{Lubensky,Renn,Nieuwenhuizen} in the sense that the asymptotic behavior of the survival probability
has been determined analytically. The one-dimensional case can be solved in the
sense that the survival probability can be computed exactly by the standard tools of
random walk theory~\cite{Weiss,Luck}. In the meanwhile the mobile traps case has resisted
analytic approaches, even in one space dimension, and it is only very recently that
extensive numerical studies were devoted to unravelling its fine
properties~\cite{grassberger}. What can be inferred, however, from 
phenomenological
arguments~\cite{rednerkang} or from low trap density expansions~\cite{szabo},
is that the survival probability of the tagged walker decays with time $S$ as
\begin{equation}
Z(S)\sim\left\{
\begin{array}{ll}
\exp(-c_d\rho S^{d/2})&d<2\\
\exp(-c_d \rho S)&d>2
\end{array}\right.
\end{equation}
where $c_d$ is a dimension dependent constant (universal in $d<2$, but non
universal in $d>2$). What makes the problem so analytically difficult to
tackle? One answer is that we are dealing with a truly
dynamical problem. There is no underlying static partition function, as opposed
to the static traps situation. This prevents in particular, the use of quantum field theory
methods which proved so successful for static traps. Of course other types of field-theoretic mappings exist
(dynamical theories built after the Doi-Peliti mapping), but
all of them fail~\cite{FvWKO} to predict the asymptotic behaviour of the survival
probability: just as is the case for static traps, this is a strong coupling problem not accessible with the usual
perturbative toolbox. Finally, it is interesting to note the close connection
between the present trapping problem and the study of systems of vicious
walkers~\cite{vicious} which exhibit some analytic similarities (the one trap problem is
the two vicious walker problem).\\

Our goal in this article is to cast the problem in a form suitable for a
systematic analytic treatment. We will go beyond the existing results for the
survival probability by setting our calculation in a systematic framework and by
giving amplitudes to leading order in an $\eps=2-d$ expansion.

\subsection{Notations}
Consider $N$ random walkers in a volume $V=L^d$ and denote by $\rho=\frac{N}{V}$
their average density (eventually the thermodynamic limit $N,V\to\infty$ with
fixed $\rho$ shall be taken). The positions of those walkers are denoted by $\bx_i(s)$,
their
initial positions $\bx_i(0)$ being random in space. Let $\br(s)$ denote the position of
the tagged walker, which starts from the origin at time 0. We employ Brownian
motion functionals to describe the dynamics of the system. The action governing
the dynamics of the set of $N+1$ walkers is $A_0+A_{\text{int}}$, where
\begin{equation}
A_0[\br,\{\bx_i\}]=\frac{1}{2}\int_0^S\dd s\left[\left(\frac{\dd \br}{\dd
s}\right)^2+\sum_i\left(\frac{\dd \bx_i}{\dd s}\right)^2\right]
\end{equation}
encodes the free motion of the particles. The interaction term reads
\begin{equation}
A_{\text{int}}[\br,\{\bx_i\}]=\beta\sum_i\int_0^S\dd
s\;\delta^{(d)}(\br(s)-\bx_i(s))
\end{equation}
It describes the trapping of $\br$ by the mobile traps $\bx_i$'s at a rate $\beta$.\\

We define two types of averages, that with respect to the free action with an
index 0,
\begin{equation}
\langle ...\rangle_0=\int{\cal D}\br\delta^{(d)}(\br(0))\prod_i\int{\cal
D}\bx_i\int\frac{\dd^d x_i(0)}{V}...\ee^{-A_0}
\end{equation} 
and that with respect to the full interacting process,
\begin{equation}
\langle ...\rangle=\langle...\ee^{-A_{\text{int}}}\rangle_0
\end{equation}
without any index. Normalization is chosen so that  $\langle 1\rangle_0=1$, so that $\langle 1\rangle<1$.
\subsection{Quantities of interest}
We define the survival probability after time $S$ by
\begin{equation}
Z(S)\equiv \langle 1\rangle
\end{equation}
and introduce the function $Z(\bq,S)$ defined by
\begin{equation}
Z(\bq,S)\equiv \langle \ee^{i\bq.\br(S)}\rangle
\end{equation}
which allows to define the Fourier transform of the probability of presence a
walk conditioned to survive up until time $S$:
\begin{equation}
G(\bq,S)=\frac{Z(\bq,S)}{Z(S)}
\end{equation}
From the knowledge of $G(\bq,S)$ it should eventually be possible to deduce the mean square
displacement $R^2\equiv-2d\frac{\p G}{\p q^2}$ of a walk conditioned to
survive. 
It is not {\it a priori} obvious
whether anomalous diffusion should occur and investigating this issue is beyond
the scope of the present work.

\subsection{Preliminary analysis}
\subsubsection{Mobile traps, immobile tagged particle}
Here we give the exact result for the survival probability: denote by
$f(\bx,\tau)$ the probability that a random walk starting from $\bx$ reaches the
origin for the first time exactly at time $\tau$, the survival probability of
tagged particle (located at the origin) is given by
\begin{equation}\label{marcheurstatique}
Z(S)=V^{-N}\sum_{\{\bx_i\}}\prod_{i=1}^N\left(1-\sum_{\tau=0}^{S}f(\bx_i,\tau)\right)
=(1-\frac{1}{V}\sum_{\bx}\sum_\tau f(\bx,\tau))^N\stackrel{N\to\infty}{\simeq}\ee^{-\rho
{\cal N}(S)}
\end{equation}
where ${\cal N}(S)$ ($\simeq
\frac{S^{d/2}(2\pi)^{d/2}}{\Gamma(\frac{\eps}{2})\Gamma(2-\frac{\eps}{2})}$ if $d<2$ and $\simeq $ if $d>2$) is the average of the number of distinct sites visited by a
random walk after $\tau$ steps (a functional derivation of this century old
lattice result can be deduced from \cite{Duplantier} : ${\cal N}(S)=\int\dd^d
x\langle(1-\ee^{-\beta\int\dd s\delta^{(d)}(\br(s)-\bx)})\rangle_0=\beta
S\sum_{n\geq 0}\frac{(-1)^n(\beta
S^{\eps/2}(2\pi)^{-d/2}\Gamma(\eps/2))^n}{\Gamma(2+n\frac{\eps}{2})}$). This
already reproduces the characteristic
features of the conjectured behaviour. Note that the result
(\ref{marcheurstatique}) is exact --a property already noticed in
\cite{blumenzumofenklafter}-- as long
as the tagged walker is immobile.
\subsubsection{Dimensionless variables}
It is interesting to note that $d=2$ appears to be the upper critical dimension in this
problem. We shall henceforth set $\eps\equiv 2-d$. This can be seen by scaling out the
walk's length $S$. We build two independent dimensionless couplings
\begin{equation}
u\equiv\beta \rho S,\;\;\;v=(4\pi)^{-d/2}\rho^{-1}S^{-d/2}
\end{equation}
and further define
\begin{equation}
w\equiv uv= (4\pi)^{-d/2}\beta S^{\frac{\eps}{2}}
\end{equation}
The local trap density,
\begin{equation}
\rho(\bx,t)\equiv\sum_{i=1}^N\delta^{(d)}(\bx-\bx_i(t))
\end{equation}
is a Poissonian variable and it is therefore difficult to exactly integrate out its
fluctuations. Note that diffusion noise is often well-described by a Gaussian
white noise. Such an approximation allows density to become negative (if extremely low
probability), and since those very regions of low trap density govern to
dynamics, the importance of those unphysical configurations is enhanced which
completely messes any analysis based on the Gaussian noise approximation.  We are going to attempt an expansion in powers of the trapping probability
(which is proportional to $\beta$) and
eventually take the limit in which this probability goes to 1. The trapping rate
$\beta$ can be viewed as an extra parameter that can be used to probe the
model's properties. 
\section{Pertubation expansion for $Z(S)$}
\subsection{General structure of the expansion}
We want to evaluate $Z(\bq,S)=\langle\ee^{i\bq.\br(S)}\rangle$. First we expand
$\ee^{-A_{\text{int}}}$ in powers of the trapping rate $\beta$, which yields
\begin{equation}\label{expansion}\begin{split}
\ee^{-A_{\text{int}}}=&\sum_{n=0}^{+\infty}\frac{(-\beta)^n}{n!}\int_0^S\dd s_1...\dd
s_n\int\frac{\dd^d k_1}{(2\pi)^d}...\frac{\dd^d
k_n}{(2\pi)^d}\sum_{i_1,...,i_n=1}^N\\&
\exp\left[{i\int_0^S\dd\tau
\by(\tau)\sum_{j=1}^n\bk_j\Theta(s_j-\tau)}\right]\\&\times
\exp\left[-{i\sum_{j=1}^n\int_0^S\dd\tau\by_{i_j}(\tau).\bk_j\Theta(s_j-\tau)}\right]\\
&\times \exp\left[-i\sum_{j=1}^n\bx_{i_j}(0).\bk_j\right]
\end{split}\end{equation}
We have denoted by $\by(s)=\frac{\dd \br}{\dd s}$ the velocity of the tagged
walker and by $\by_i=\frac{\dd \bx_i}{\dd s}$ that
of trap $i$. Consider the $n^{\text{th}}$ order term. It involves an average over $n$ random
walkers $\bx_{i_1}$,...,$\bx_{i_n}$. We depict a general $n^{\text{th}}$ order term by the follwoing diagram:
\begin{figure}
$$\input{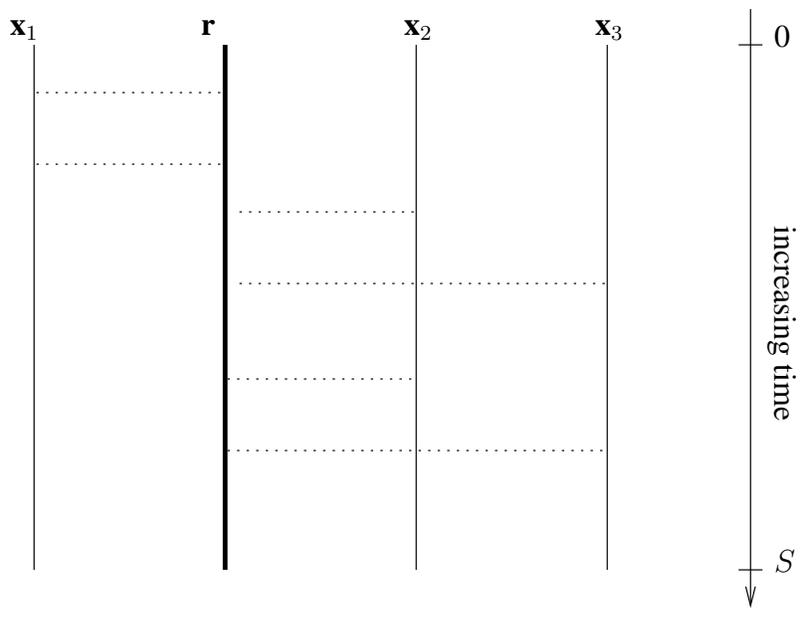}$$
\caption{The bold trajectory of the tagged walker $\br$ meets that of the traps
$\bx_{1}$, $\bx_{2}$, $\bx_{3}$. The depicted diagram is of order $6$ (there are six
intersection times).}
\end{figure}
In a general diagram, the number of horizontal lines stands for the number of
intersections that have take place between times 0 and $S$, this is also the power of $\beta$ in the
perturbation expansion. The number of vertical lines, at order $n$, stands for
the number of {\it distinct} traps that will intersect the tagged walkers's
trajectory. A diagram of order $n$ involving $m\leq n$ distinct traps will be
proportional to $u^{n} v^{m}=u^{m} w^{n-m}$. It is not a trivial task to determine the proportionality constant; this will
depend in a complicated manner on the ``topology'' of the diagram (the latter constant
will be a function of $\eps$).
\subsection{An example: diagrams of order 4}
As an example, we explicitly compute the contributions arising from the term
$n=4$ in the expansion (\ref{expansion}). We depict those contributions by a
series of diagrams:
\begin{figure}
$$\input{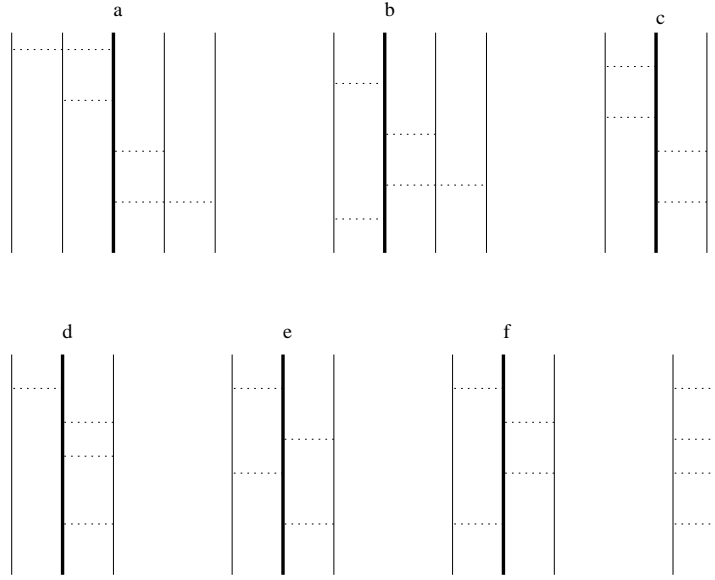}$$
\caption{Shown are diagrams of order 4. Diagram (a) is proportional to $u^4$,
diagram (b) to $u^3 w$, diagrams (c,d,e,f) to $u^2w^2$ and diagram (g) to $uw^3$.}
\end{figure}
We now state the explicit evaluation from which one can induct the general
properties of those diagrams, such as their leading UV divergence. Diagrams
$(a,b,c,d,g)$ exhibit no overlapping loops.
\begin{equation}\begin{split}
(a)=&\frac{u^4}{4!}
\end{split}\end{equation}
\begin{equation}\begin{split}
(b)=&\frac{1}{2}u^3w\Gamma(\eps/2)\frac{1}{\Gamma(2+\eps/2)}
\end{split}\end{equation}
\begin{equation}\begin{split}
(c)=&\frac 12 u^2\left({w}{\Gamma(\eps/2)}\right)^2\frac{2}{\Gamma(3+\eps)}
\end{split}\end{equation}
\begin{equation}\begin{split}
(d)=&u^2\left({w}{\Gamma(\eps/2)}\right)^2\frac{1}{\Gamma(2+\eps)}
\end{split}\end{equation}
Diagrams $(e,f)$ possess an overlapping loop:
\begin{equation}\begin{split}
(e)=&u^2w^2\int_{0<x_1<x_3<x_2<x_4<1}\dd x_1\dd x_2\dd x_3\dd
x_4\;((x_4-x_3)(x_2-x_1)-\frac{1}{4}(x_2-x_3)^{2})^{-\frac{d}{2}} \\=&\int\dd
t_1\dd t_2\dd t_3\Theta(1-t_1-t_2-t_3)(1-t_1-t_2-t_3)(t_1 t_2+t_2 t_3+t_3
t_1+\frac{3}{4}t_2^2)^{-\frac{d}{2}}\\ 
=&\frac{1}{\eps}\left(-\frac{50}{3}-48\ln
3+\frac{824}{9}\ln 2-\frac{4}{\sqrt{3}}\ln\frac{2-\sqrt{3}}{2+\sqrt{3}}\right)+{\cal O}(1)
\end{split}\end{equation}
The coefficient $\frac{3}{4}=\frac{4D(D+1)}{(2D+1)^2}$, where $D=\frac 12$ is the trap
diffusion constant. If the traps were static, one would have $D=0$ and one would
recover a standard $\phi^4$ propagator diagram with a double pole in $4-d$. A nonzero
diffusion constant acts as a partial UV regulator which shifts the existence of
a UV
divergence from $d=4$ down to $d=2$.   
\begin{equation}\begin{split}
(f)=&u^2 w^2\int_{0<x_1<x_3<x_4<x_2<1}\dd x_1\dd x_2\dd x_3\dd x_4\;
(x_4-x_3)^{-\frac{d}{2}}(x_2-x_1-\frac{1}{4}(x_4-x_3))^{-\frac{d}{2}}\\
=&u^2
w^2\int\dd t_1\dd t_2 \dd t_3\Theta(1-t_1-t_2-t_3)(1-t_1-t_2-t_3)
t_2^{-\frac{d}{2}}(t_1 + \frac{3}{4}t_2 + t_3)^{-\frac{d}{2}}
\\=&u^2 w^2\left(\frac{1}{\eps}+{\cal
O}(1)\right)
\end{split}\end{equation}
Finally,
\begin{equation}\begin{split}
(g)=&u\left({w}{\Gamma(\frac{\eps}{2})}\right)^3\frac{1}{\Gamma(2+3\frac{\eps}{2})}
\end{split}\end{equation}
The conclusion to be drawn from those explicit expression is the way the
short-distance divergences (the $\eps$ poles) are interwined with the topology of the diagrams
which are considered.
\subsection{Nonoverlapping diagrams}
We now focus the subclass of nonoverlapping diagrams: in those diagrams, there
are no overlapping loops. In order to fully characterize such a diagram, one
needs:
\begin{itemize}
\item its order, denoted by $M$; this is also the total number of intersections.
\item the number $n$ of distinct traps intersecting the walker's path.
\item the number of times $m_j$ trap $j$ ($j=1,...,n$) intersects (in a row) the
walker's path. Of course, $\sum_{j=1}^n m_j=M$. 
\item among the set $\{m_j\}_{j=1}^{n}$, we count the number of $m_j$'s which
are equal. If they take $\nu$ distinct values, we call $p_\ell$,
$\ell=1,...,\nu$, the number of $m_j$ coefficients that take the same value indexed by
$\ell$. This is necessary to properly determine the symmetry coefficient of each
diagram. The allowed values of $\nu$ run from 1 through $\nu_n=n$ if $n$ can be written in the form $n=k(k+1)/2$ ($k$ an integer)
and $\nu_n=n-1$ otherwise.
\end{itemize}
In principle the following calculation holds for $m_j\geq 2$ but it can be seen
to trivially extend to $m_j=1$ without modification.\\

First consider a particular diagram in which the $M$ intersection times ordered
\begin{equation}
0\leq s^{(1)}_1\leq s^{(1)}_2\leq...\leq s_{m_1}^{(1)}\leq s_1^{(2)}\leq...\leq s_{m_n}^{(n)}\leq S 
\end{equation}
where the upper index $j$ ($j=1,..,n$) denotes the trap that intersects the walker's path and the
lower index $i$ counts the number of intersections of trap $j$ ($i=1...,m_j$).
Its graphical representation is as follows:
\begin{figure}
$$\input{art8fig4.pstex_t}$$
\end{figure}
We call $D_{M,n}$ the value of that diagram. Given that sequence of intersections, we have to evaluate
\begin{equation}\begin{split}
D_{M,n}=&\frac{(-\beta)^M}{M!}N(N-1)..(N-n+1)
\int\dd s^{(1)}_1...\dd s_{m_n}^{(n)}
\int\frac{\dd^d k_1^{(1)}}{(2\pi)^d}...
\frac{\dd^d k_{m_n}^{(n)}}{(2\pi)^d}\\&
\langle\exp\left[i\int_0^S\dd\tau
\by(\tau)\left(\sum_{j=1}^n\sum_{i=1}^{m_j}\bk^{(j)}_i\Theta(s_i^{(j)}-\tau)\right)\right]\\&
\exp\left[-i\int_0^S\dd\tau
\left(\sum_{j=1}^n\by_j(\tau)\sum_{i=1}^{m_j}\bk^{(j)}_i\Theta(s_i^{(j)}-\tau)\right)\right]\\&
\exp\left[-i\sum_{j=1}^n \bx_j(0)
\sum_{i=1}^{m_j}\bk_i^{(j)}\right]
\rangle_0\\
=&\frac{(-1)^M}{M!}
\frac{N(N-1)...(N-n+1)}{M^n}u^n w^{M-n}\\&
\times\int_0^1\dd t_1\int_0^{1-t_1}\dd t_2...\int_0^{1-t_1...-t_{M-1}}\dd
t_M\\&\left(t_1^{0}\;t_2^{-\frac{d}{2}}\;t_3^{-\frac{d}{2}}\;...\;t^0_{m_1+1}\;
t^{-\frac{d}{2}}_{m_1+2}\;...\;t^{0}_{m_1+...+m_{n-1}+1}\;...\;t_M^{-\frac{d}{2}}\right)
\\
=&\frac{(-1)^M}{M!}
\frac{N(N-1)..(N-n+1)}{N^n}
\frac{u^n (w\Gamma(\frac{\eps}{2}))^{M-n}}{\Gamma(n+1+(M-n)\frac{\eps}{2})}
\end{split}\end{equation}
While the explicit evaluation is not a trivial task, the final result is
particularly simple.\\

We now consider a general diagram of order $M$ with $n$ intersecting traps
characterized by $\{m_j\}_{j=1,...,n}, \{p_\ell\}_{\ell=1,...,\nu}$. We do not
specify the order in which the traps intersect the walker's path. Such a diagram
has a value
\begin{equation}
\frac{M!}{\prod_{j=1}^n m_j!\prod_{\ell=1}^\nu p_\ell!}\;n!\;\left(\prod_{j=1}^n
m_j!\right)\;D_{M,n}=(-1)^M\frac{ n!}{\prod_{\ell=1}^\nu p_\ell!}
\frac{u^n (w\Gamma(\frac{\eps}{2}))^{M-n}}{\Gamma(n+1+(M-n)\frac{\eps}{2})}
\end{equation}
At a given order $M$, a diagram
involving $n$ distinct walkers has a higher $\eps$-divergence if it is free of
overlapping loops. Retaining for each order $M$ only the nonoverlapping diagrams,
we get the survival probabiilty 
\begin{equation}
Z(S)=1+\sum_{M=1}^{+\infty}\sum_{n=1}^{M}\sum_{m_1+...+m_n=M}
\sum_{\nu=1}^M\sum_{p_1+...+p_\nu=n}(-1)^M\frac{ n!}{\prod_{\ell=1}^\nu p_\ell!}
\frac{u^n (w\Gamma(\frac{\eps}{2}))^{M-n}}{\Gamma(n+1+(M-n)\frac{\eps}{2})}
\end{equation}
Since we are concerned with leading divergences, we will
approximate 
\begin{equation}
\frac{1}{\Gamma(n+1+(M-n)\frac{\eps}{2})}\simeq\frac{1}{n!}
\end{equation}
Summing over all values of $M-n$ yields a factor
$(1+\frac{w}{\Gamma(\frac{\eps}{2})})^{-n}$
Setting $g\equiv \frac{u}{1+{w}{\Gamma(\frac{\eps}{2})}}$, we are left with
\begin{equation}
Z(S)=\ee^{-g}
\end{equation}
In the limit $u,w\to\infty$ with $\frac{u}{w}$ fixed, this leads to the asymptotic behaviour
\begin{equation}\label{beau}
Z(S)\sim \ee^{-c_\eps \rho S^{d/2}}
\end{equation}
with $c_\eps=2\pi{\eps}+{\cal O}(\eps^2)$. We emphasize that this is not a low
density expansion since the variable $\frac{u}{w}\propto \rho S^{d/2}$ is held
constant. We could also have written
\begin{equation}
Z(S)\sim \exp\left[-u\sum_{n=0}^{\infty}\frac{(-w\Gamma(\frac{\eps}{2}))^n}{\Gamma(2+n\frac{\eps}{2})}\right]
\end{equation}
which is, to leading order in $\eps$, equivalent to the previously mentioned result.\\

The only approximation made is to neglect overlapping
diagrams. This is equivalent to performing an $\eps$-expansion of $c_\eps$.
Compare two diagrams that differ by the presence of an overlapping loop
instead of a non-overlapping loop: in the latter, the leading short distance
singularity is not constrained by time ordering. In the former, the time
ordering associated with an overlapping loop acts as a partial short-distance
regulator. The related UV divergence will always be smoother than in the
nonoverlapping counterpart. However the
power of $S$ that appears in Eq.~(\ref{beau}) is exact since, at a given order in $u$ and $w$,
diagrams with or without overlapping loops have the same $S$ dependence.
We argue that a
diagram containing at least a pair of overlapping loops has a softer $\eps$
divergence than the equivalent diagram in which the overlapping loops have been
disentangled. 
\subsection{Perturbation expansion of $Z(\bq,S)$}
As before, we shall now keep in the perturbation expansion only diagrams
containing no loop overlap, this leads to
\begin{equation}
\begin{split}
D_{M,n}=&\frac{(-\beta)^M}{M!}N(N-1)..(N-n+1)
\int\dd s^{(1)}_1...\dd s_{m_n}^{(n)}
\int\frac{\dd^d k_1^{(1)}}{(2\pi)^d}...
\frac{\dd^d k_{m_n}^{(n)}}{(2\pi)^d}\\&
\langle\exp\left[i\int_0^S\dd\tau
\by(\tau)\left(\bq\Theta(S-\tau)+\sum_{j=1}^n\sum_{i=1}^{m_j}\bk^{(j)}_i\Theta(s_i^{(j)}-\tau)\right)\right]\\&
\exp\left[-i\int_0^S\dd\tau
\left(\sum_{j=1}^n\by_j(\tau)\sum_{i=1}^{m_j}\bk^{(j)}_i\Theta(s_i^{(j)}-\tau)\right)\right]\\&
\exp\left[-i\sum_{j=1}^n \bx_j(0)
\sum_{i=1}^{m_j}\bk_i^{(j)}\right]
\rangle_0\\
=&\frac{(-1)^M}{M!}\ee^{-\frac{q^2S}{2}}
\frac{N(N-1)...(N-n+1)}{M^n}u^n w^{M-n}\\&
\times\int_0^1\dd t_1\int_0^{1-t_1}\dd t_2...\int_0^{1-t_1...-t_{M-1}}\dd
t_M\\&\left(t_1^{0}\;t_2^{-\frac{d}{2}}\ee^{\frac{q^2St_2}{4}}\;t_3^{\frac{d}{2}}\ee^{\frac{q^2St_3}{4}}\;...\;t^0_{m_1+1}\;
t^{-\frac{d}{2}}_{m_1+2}\ee^{\frac{q^2St_{m_1+2}}{4}}\;...\;t^{0}_{m_1+...+m_{n-1}+1}\;...\;t_M^{-\frac{d}{2}}\ee^{\frac{q^2St_M}{4}}\right)
\\
=&\ee^{-\frac{q^2 S}{4}}\frac{(-1)^M}{M!}
u^n (w\Gamma(\frac{\eps}{2}))^{M-n}\int\frac{\dd z}{2\pi
i}\ee^{z}\frac{1}{(z+\frac{q^2S}{4})^{n+1}}z^{-(M-n)\frac{\eps}{2}}
\end{split}
\end{equation}
where the integration path runs from $-i\infty+0$ up to $+i\infty+0$. The final
result now reads 
\begin{equation}
Z(\bq,S)=\ee^{-\frac{q^2 S}{4}}\int\frac{\dd z}{2\pi i}\ee^z\int_0^{+\infty}\dd
x\ee^{-x(z+\frac{q^2S}{4})}\exp\left[-\frac{u
x}{1+\frac{w}{\Gamma(\eps/2)z^{\eps/2}}}\right]
\end{equation}
so that $G(\bq,S)=\ee^{-\frac{q^2 S}{2}}$. Diffusion is not affected at this
level of the approximation.
\section{Discussion}
\subsection{Comparison with the static trap problem}
When traps are static, the Brownian motion functional depends on the tagged
walkers trajectory only through its local time $\chi(\bx,S)\equiv \int_0^S\dd
s\;\delta^{(d)}(\br(s)-\bx)$, hence it can be mapped onto an $O(n)$ field theory with an
interaction term than can be shown to be
\begin{equation}A_{\text{eff}}[\br]=-\int\dd^d x\;\ln\left(1-\rho+\rho\ee^{-\beta\int_0^S\dd s \delta^{(d)}(\br(s)-\bx)}\right)\to
-\int\dd^d x\;\ln\left(1-\rho+\rho\ee^{-\beta\phi^2}\right)
\end{equation}
We refer the reader to \cite{Lubensky, Renn, Nieuwenhuizen} and references therein for
details on how to derive this correspondence and how to exploit it. For small values of $\beta$
or of the field
this produces an attractive $-\phi^4$ theory (which has no minimum, and which is
equivalent to truncating the walker's effective self-interation potential to that of a
self-attracting walk $-\beta^2\rho^2\int\dd s \dd
s'\;\delta^{(d)}(\br(s)-\br(s'))$), hence the complete knowledge of
the interaction potential is necessary. No truncation of the interaction
potential can be used.\\

Attempting an expansion in powers of $\beta$ will generate self-interaction diagrams that
must be evaluated in a standard way (a loop corresponds to a $\delta(\br(s)-\br(s'))$
interaction). Loops in diagrams occur when a trap is visited at least twice. All those
features have their counterpart when traps are mobile.

If traps are mobile, the best Langevin equation that describes the density of the diffusing traps is
\begin{equation}
\p_t\rho-D\Delta\rho=\eta,\;\;\;
\langle\eta(\bx,t)\eta(\bx',t')\rangle=2D\rho\nabla\nabla'\delta(t-t')\delta^{(d)}(\bx-\bx')
\end{equation}
If this Langevin equation were exact (which it is not since the density is a Poissonian
variable and not a Gaussian one), this would lead to an effective interaction term
depending  only upon the path $\br$
\begin{equation}
A_{\text{eff}}[\br]=-\beta^2\rho^2\int_0^S\dd s\dd
s'\frac{1}{(4\pi D|s-s'|)^{d/2}}\exp\left[-\frac{(\br(s)-\br(s'))^2}{4D|s-s'|}\right]
\end{equation}
We now see that the $\delta^{(d)}(\br(s)-\br(s'))$-function appearing in the static traps case has been
replaced by $g(\br(s)-\br(s'))$ (with $g(\bx,\tau)\equiv (4\pi D|\tau|)^{-d/2}\ee^{-\bx^2/(4D|\tau|)}$
and $D=1/2$). In fact, any given graph of the static traps problem has its mobile traps
counterpart, which is evaluated simply by replacing the $\delta$-function in the loops
(whether overlapping or not)
by
the traps propagator $g$. There is a one-to-one correspondence between the graphs of the
two problems.
\subsection{Beyond the leading order}
For static traps the mean-square displacement is known~\cite{Weiss} to scale as $R^2\sim
S^{\frac{d+1}{d+2}}$ and the return to the origin probability as
$S^{-\frac{d}{d+2}}$. Within the framework of our approximation, if traps are
diffusing, we find that both the mean-square displacement and the return to the
origin probability take their simple random walk asymptotics. Expanding the
function $G(\bq,S)$ beyond leading order in $\eps$ shows that this picture might
not hold for the full interacting theory. Indeed, the neglected UV divergencies
 could be responsible for a modified scaling behavior. For instance, one can
show that $G(\bq,S)=\ee^{-\frac{q^2 S}{2}}\left(1+\frac{u^4
v^2}{\eps}F(q^2S)+{\cal O}(1)\right)$, with $F$ a known function that can easily
be expanded
in the vicinity of 0. Deeper insight into the structure of the
perturbation expansion is required before
exploiting such results in the spirit of the approach developped by Duplantier
for polymers~\cite{Bertrand}. 

\subsection{Conclusion}
We have given the first systematic approximation scheme for the survival
probability of a random walk evolving in a medium infested with freely diffusing
traps. Our method exploits the properties of low-dimensional Brownian motion. As
opposed to the static trap case, no mapping to an equilibrium-like field theory
exists and due to the strong coupling nature of the problem the elegant field-theoretic methods of reaction-diffusion problems
fail. Extending the Brownian motion formalism to truly nonequilibrium processes
therefore proved unexpectedly fruitful. The method which we have elaborated should now be put to the test on a series of
related problems which we now list. The first question
that comes to mind is concerned with the scaling properties of a surviving path:
what is the scaling behaviour of the walker's mean-square displacement between
0 and $S$, provided it has survived up until that time. Questions of lesser
importance
concern the number of distinct sites visited by the walker, its return to the
origin probability or the algebraic area swept (still restraining the average to
surviving paths). Numerical approaches are usually difficult to
implement~\cite{grassberger} when
it comes to exploring those
properties due to the poor statistics over surviving events.  Future work should address those
issues.\\

\noindent {\bf Acknowledgments :} The author would like to thank Henk
Hilhorst and Bertrand Duplantier for repeated discussions on this work.

\end{document}